\documentclass[twocolumn,aps,prl,english,twocolumn,epsfig,groupedaddress]{revtex4}
\usepackage{epsfig}
\usepackage{amsmath}
\usepackage[latin1]{inputenc}
\usepackage{latexsym}
\usepackage{graphicx}
\begin{document}
\title{Effects of correlated disorder on the magnetism of double exchange systems}
\author{G. Bouzerar$^{1,2}$ and O. C\'epas$^{3}$}
\affiliation{
$1.$ Institut N\'eel, d\'epartement MCBT, 25 avenue des Martyrs, C.N.R.S., B.P. 166 38042 Grenoble Cedex 09, France \\
$2.$ Institut Laue Langevin, BP156, 38042 Grenoble Cedex, France. \\
$3.$ Laboratoire de physique th\'eorique de la mati\`ere condens\'ee, C.N.R.S. UMR 7600, Universit\'e Pierre-et-Marie-Curie, Paris, France.}

\date{\today}

\begin{abstract}

We study the effects of short-range correlated disorder arising from
chemical dopants or local lattice distortions, on the ferromagnetism
of 3d double exchange systems. For this, we integrate out the
carriers and treat the resulting disordered spin Hamiltonian within
local random phase approximation, whose reliability is shown by direct
comparison with Monte Carlo simulations. We find large scale
inhomogeneities in the charge, couplings and spin densities. Compared
with the homogeneous case, we obtain larger Curie temperatures
($T_{C}$) and very small spin stiffnesses ($D$). As a result, large
variations of $\frac{D}{T_{C}}$ measured in manganites may be
explained by correlated disorder.  This work also provides a
microscopic model for Griffiths phases in double exchange systems.

\end{abstract}
\maketitle

The interest for disordered magnetic systems such as thin magnetic
films of transition metal alloys (Fe-Ni,Co-Ni,...), diluted magnetic
semiconductors (Ga$_{1-x}$Mn$_{x}$As$, $Ge$_{1-x}$Mn$_{x}$,...), d$^0$ materials (HfO$_2$,CaO,..) or manganites
(Re$_x$A$_{1-x}$MnO$_3$, where Re is a rare-earth ion and A an
alkaline ion) has considerably increased during this last decade. One
of the reasons is the potential of some of the materials to be
incorporated in technological devices. Some of them play a very special role: systems which contain large scale
inhomogeneities. Inhomogeneities can appear during the growth of the
sample by molecular beam epitaxy for example but can also
result from the interplay between many degrees of
freedom (charge, spin, orbital, phonons). This is for example the
case in manganites. It is known that manganites are strongly
inhomogeneous at the nanometer scale: (i) large-scale structures in
the \textit{charge} density were seen by electron diffraction of thin films
\cite{Uehara}, or tunneling spectroscopy \cite{Fath}; (ii) evidence
for inhomogeneous \textit{spin} density was found in neutron diffuse
scattering \cite{Moussadiffraction}, or NMR \cite{Papav}; (iii)
localized spin waves also suggest the presence of confining potentials
\cite{Hennion}. There are also clear evidences of inhomogeneous
structures above $T_{C}$, which were interpreted
\cite{Salamon,Deisenhofer} as a Griffiths phase \cite{Bray}.  Their
microscopic origin is one of the central issues of
the physics of manganites; it includes phase separation frustrated by
long-range Coulomb interaction \cite{Dagotto}, chemical disorder
\cite{deGennes,Nagaev}, polarons \cite{Rama,Shenoy}.

In this paper, we argue that the way the disorder is modelled is
important to understand large-scale inhomogeneous structures in 3d and
to explain the Griffiths phase \cite{Salamon}.  For this we consider a
model where the disorder is \textit{correlated} at short
distances. This model gives a possible explanation for the broad and
multi-modal distribution of NMR lines
\cite{Papav}, or the wide distribution of Curie temperatures $T_C$
\cite{Hwang} and spin stiffnesses \cite{Tapan0} measured in different
materials for the same carrier density. The microscopic origin of the
correlated disorder could be chemical or polaronic. For instance, in
Re$_{1-x}$A$_{x}$MnO$_{3}$ the dopant A$^{2+}$ which substitutes
Re$^{3+}$ creates a strong Coulomb potential on its neighborhood and
in particular on the eight nearest neighbor Mn sites surrounding it
\cite{Singh}. This is the model of ``color-centers'' initially
discussed by de Gennes \cite{deGennes}. Alternatively, local
Jahn-Teller distortions can also be seen as a source of correlated
disorder through ``cooperative phonons'', which can be mapped onto the
same model.

The 3d correlated disordered double exchange Hamiltonian we consider
reads,
\begin{eqnarray}
H=\sum_{ij\sigma} (t_{ij}c^{\dagger}_{i\sigma}c_{j\sigma}+h.c.) - J_{H}\sum_{i} \vec{S}_{i}\cdot \vec{s}_{i} + \sum_{i} \epsilon_{i}n_{i}
\label{Hamiltonian}
\end{eqnarray}
where $t_{ij}=-t$ for nearest neighbors only, $\vec{S}_{i}$ is a
classical spin localized at site $i$ ($|\vec{S}_{i}|=1$) and
$\vec{s}_i=c^{\dagger}_{i\alpha}(\vec{\sigma})_{\alpha \beta}
c_{i\beta}$, $J_{H}$ is the Hund coupling which is set to be $\infty$.
The on-site potentials $\epsilon_{i}$ may correspond, in particular,
to the chemical substitution of Re$^{3+}$ by A$^{2+}$ defined by
$\epsilon_{i}=\epsilon_D \sum_{l} x^{i}_{l}$; where the sum runs over
the $l$ nearest neighbour cations of the Mn site $i$ ($l=1,\dots,8$),
and $\epsilon_D$ is the strength of the disorder. We choose randomly
$x$ cationic sites for $A$ for which $x^{i}_{l}=1$ (otherwise
$x^{i}_{l}=0$). We emphasize that the disorder is correlated because
one dopant affects simultaneously the 8 nearest neighbour Mn
sites. With these definitions $\epsilon_{i}$ takes the discrete values
0, $\epsilon_D$, 2$\epsilon_D$,...,8$\epsilon_D$. From stoechiometry,
we would expect the hole density, $n_{h}$ to be equal to $x$, but in
order to include the local Jahn-Teller distortion picture as well, we
allow them to be different.

\begin{figure}[tbp]
\vspace{-0.3cm}
\centerline{
 \psfig{file=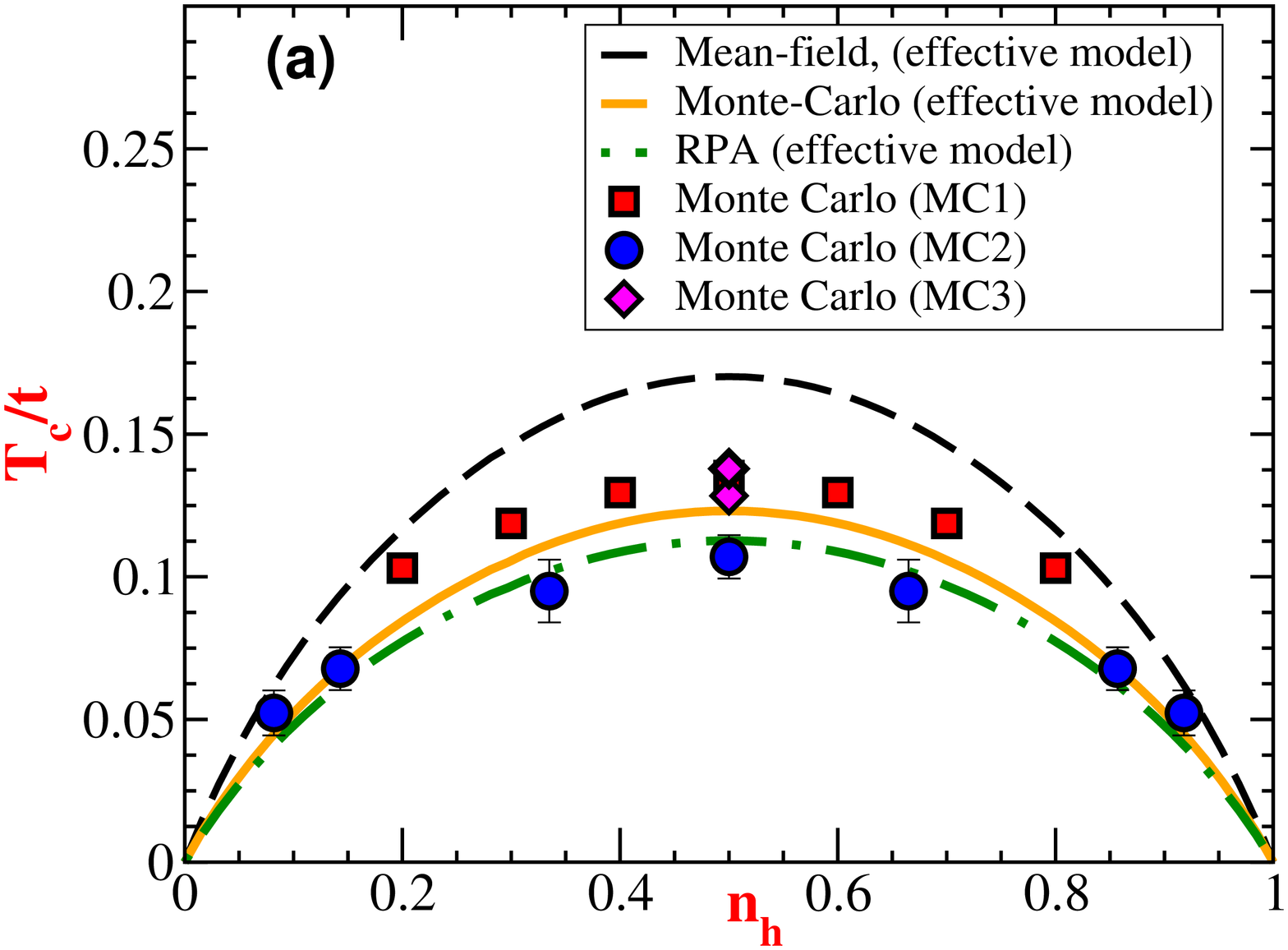,width=5.5cm,angle=-90}}
\centerline{}
\centerline{}
\vspace{-1.2cm}
\centerline{\psfig{file=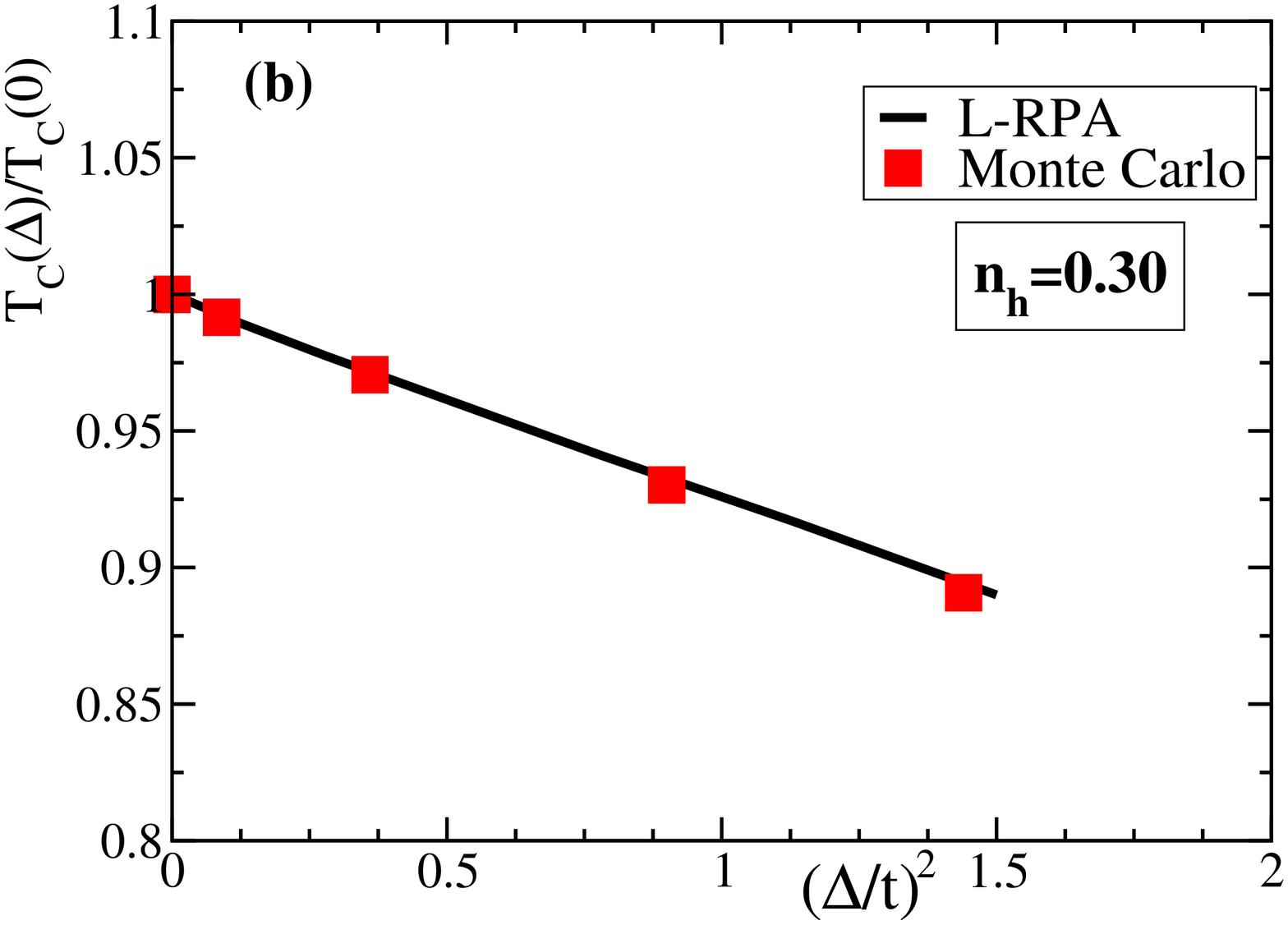,width=5cm,angle=-90}}
\vspace{0.0cm}
\caption{(Color online) (a) T$_{C}$ as a function of hole density (clean case). Lines are obtained with the effective Heisenberg Hamiltonian within Mean Field (dashed), RPA (dotted dashed) and Monte Carlo (continuous) treatments; and symbols from Monte Carlo simulations of the full double exchange model: MC1 \cite{MotomeMonteCarlo}, MC2 \cite{DagottoMonteCarlo} and MC3 \cite{Alonso}.
(b) $T_{C}$ as a function of the on-site potential width $\Delta$ for the Anderson disorder (\textit{uncorrelated}): The continuous line is obtained with Local RPA and symbols are from Monte Carlo \cite{MotomeMonteCarlo}.}
\label{Fig1}
\end{figure}

The approach we use to study this model is in two steps. First, for a
given configuration of disorder we diagonalize (\ref{Hamiltonian}) in
the real space, assuming a fully polarized ground-state at zero
temperature. This allows to define an effective Heisenberg Hamiltonian
for the classical spins, $H_{\rm eff}= \sum_{<ij>} J_{ij} \vec{S}_{i} \cdot
\vec{S}_{j}$, where the disordered couplings $\{ J_{ij} \}$ are
explicitly calculated in the limit $J_{H} \rightarrow \infty$, using
$J_{ij} = t_{ij} \langle c^{\dagger}_{i,\uparrow}
c_{j,\uparrow}\rangle /2$ \cite{Motomeexcitation,Kubo}. In the second
step, we diagonalize this Hamiltonian using the self-consistent local
random phase approximation (SC-LRPA)\cite{Bouzerar}. It consists of decoupling higher-order spin-spin Green's functions 
in the equation of motion. This introduces the local magnetizations $\langle S_i^z \rangle$, which
are self-consistently determined by using sum rules.
Spatial fluctuations due to disorder are thus treated exactly
by solving the equations numerically in real space. 
This procedure was shown to be reliable
to study dilute magnetic semiconductors where the
couplings were calculated \textit{ab-initio} \cite{Bouzerar}. SC-LRPA provides an analytical
expression for $T_C$ and allows us to study much larger systems than
those used in Monte Carlo.

In Fig.~\ref{Fig1}, we test this method by comparing $T_C$ with that of Monte Carlo
simulations for both the clean system ($\epsilon_{i}=0$) 
\cite{DagottoMonteCarlo,Calderon,Alonso,MotomeMonteCarlo} and  the
system with Anderson disorder \cite{MotomeMonteCarlo,Majumdar}. In the later case, $\epsilon_{i}$
are {\it uncorrelated} variables uniformly distributed within
$[-\frac{\Delta}{2},\frac{\Delta}{2}]$.  In Fig.~\ref{Fig1}a (clean case), the
lines are obtained by studying $H_{\rm eff}$ within a simple mean-field
theory, $T^{MF}_{C}=2J$ (dashed line), RPA, $T^{RPA}_{C} = 1.32J$
\cite{Callen} (dotted dashed line) and Monte Carlo, $T^{MC}_{C}=1.44
J$ \cite{Chen} (full line); $J=\frac{-1}{2z}\frac{\langle K
\rangle}{N}$ where the kinetic energy $\langle K \rangle$ depends on
$n_h$. For the clean system, we remind that $T_C^{RPA}=1.32J$ is obtained analytically
using $T^{RPA}_{C}=\frac{1}{3}(\sum_{\bf q} \frac{1}{E(\bf q)})^{-1}$,
where $E({\bf q})= z J(1-\gamma({\bf q}))$ is the magnon dispersion, $z$ the coordination number, and $\gamma({\bf
q})= \frac{1}{z} \sum_{\bf r_{i}} e^{i {\bf q}.{\bf r_i}}$
\cite{Callen}. This expression actually gives a very good approximation of $T_{C}$; the error compared to Monte-Carlo is $8\%$. Now,
the comparison with Monte-Carlo simulations of the full double
exchange model (symbols) shows that the difference is within $10\%$,
so that the two step approach is quantitatively reliable.  Similarly,
when Anderson disorder is added, we have found that SC-LRPA gives an
excellent agreement with Monte Carlo data (Fig.~\ref{Fig1}b),
stressing that not only the thermal fluctuations are well treated but
also the spatial fluctuations due to disorder.

\begin{figure}[htbp]
\includegraphics[width=9.50cm,angle=0]{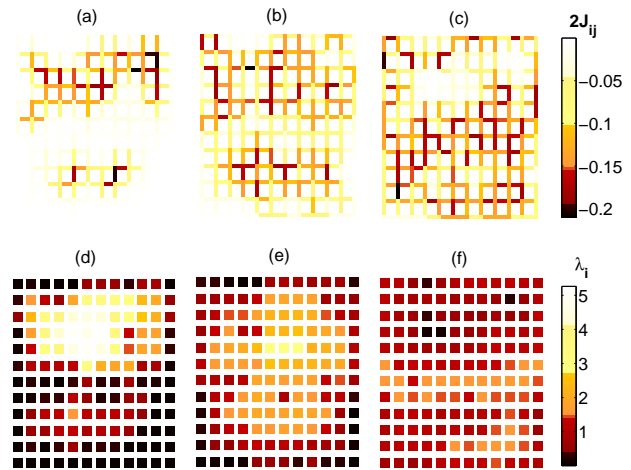}
\vspace{-0.70cm}
\caption{(Color online.) Top row: real space picture of the magnetic
couplings $2 J_{ij}$ of the effective model (\textit{correlated} disorder), on one layer of the $12^{3}$ cube. The dark
(resp. white) regions correspond to large (resp. weak)
couplings.  Bottom row: real space distribution of $\lambda_{i}=\mbox{lim}_{T \rightarrow T_{C}} \frac{\langle S_{i}^z \rangle}{m}$ on the
same layer. From left to right, $n_h=0.1$ (a),(d), $0.3$ (b),(e) and $0.5$ (c),(f).
Parameters are $x=0.3$ and $\epsilon_{D}=0.15 W$.  }
\label{Fig2}
\end{figure}

From now, we consider the model with \textit{correlated}
disorder, as discussed above. In Fig.~\ref{Fig2}a, \ref{Fig2}b,
\ref{Fig2}c, we have plotted the magnetic couplings $J_{ij}$ in a
given layer for various hole densities ($n_{h}=0.1$, 0.3 and $0.5$,
respectively). They are calculated for a fixed concentration of
randomly distributed impurities (color centers), $x=0.3$, and for a disorder
strength, $\epsilon_{D}=0.15~W$ ($W=12~t$ is the bandwidth), that is chosen to be compatible with
{\it ab initio} calculations \cite{Singh}. At \textit{low density}
(Fig.~\ref{Fig2}a), the couplings are extremely inhomogeneous in
space: we observe large clusters of strong couplings, embedded in
regions of weak couplings. The distribution function of $\{2J_{ij} \}$
(not shown) is peaked at $\approx -0.003t$ but has a very long tail up to a cutoff
of $-0.3t$ (the average is $\bar{J}=-0.02t$).
The regions of strong couplings correspond to hole rich regions with
metallic properties embedded in a hole poor matrix which is expected
to be insulating, thus leading to phase separation. This tendency will
be reinforced if antiferromagnetic superexchange couplings are taken
into account, the hole poor regions will become antiferromagnetic or
canted, as observed at very low dopings (droplets in a
canted matrix) \cite{Moussadiffraction}. For Anderson disorder, we do
not have well-defined nanoscale regions in 3d
\cite{Majumdar}, unless cooperative phonons were included
\cite{Kumar}. As the concentration of holes increases, the size of the
regions of large couplings increases, and the system becomes less
inhomogeneous. In this respect, close to half filling ($n_h=0.5$), the
nature of the disorder becomes less important, as we shall see. The
reason is that carriers with short Fermi wave length are less
sensitive to the details of the disorder. Spatial inhomogeneities in
the magnetization near $T_C$ are directly seen in the distribution
of $\lambda_{i}=\mbox{lim}_{T \rightarrow T_{C}} \frac{\langle S_{i}^z
\rangle}{m}$, where $m$ is the averaged magnetization
(Fig.~\ref{Fig2}).  For a nearly homogeneous state, $\lambda_i$ is
close to $1$, as seen in Fig.~\ref{Fig2}f.  At low densities, we see a
very inhomogeneous texture of $\lambda_i$ (Fig.~\ref{Fig2}d), with
local droplets with $\lambda_i$ as high as $\approx 4-5$, surrounded
by a region with very small local magnetizations. In this case, the
distribution of the magnetizations is multi-modal. In between
(Fig.~\ref{Fig2}b), the droplet increases in size and $\lambda_{i}
\approx 2$ is reduced with respect to Fig.~\ref{Fig2}a, the
distribution has only one broad peak. These results resemble NMR
results where multimodal distributions occur at low dopings and get
broader for higher doping \cite{Papav}.

\begin{figure}[tbp]
\vspace{-0.5cm}
\includegraphics[width=6.cm,angle=-90]{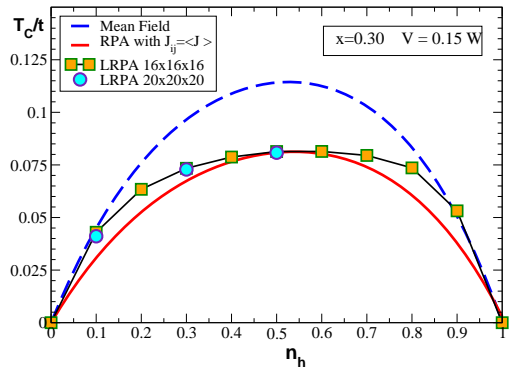}
\caption{(Color online.) $T_{C}$ (symbols) as a function of hole density for the
model with correlated disorder calculated by SC-LRPA
(averaged over 100 configurations of disorder). Also given are the
mean-field (dashed line), RPA with all couplings identical
$J_{ij}=\bar{J}$ (continuous line). Parameters are $x=0.3$ and
$\epsilon_{D}=0.15 W$. Calculations are done for sizes $16^{3}$ and $20^{3}$.  }
\label{Fig3}
\end{figure}

In Fig.~\ref{Fig3}, we give $T_{C}$ averaged over at
least 100 disorder configurations (symbols).  To see clearly the
role of the inhomogeneities, we have also indicated what $T_{C}$ would be if we replace all couplings by their
average, defined by $\bar{J}= \frac{1}{z N}\sum_{ij} J_{ij} =\langle
J_{ij} \rangle_{dis} $ (lines). The results are almost identical
for $n_{h}$ close to $0.5$ but strongly differ otherwise. Similarly we
have found (not shown) that for Anderson disorder, T$_{C}$ is also extremely close to that of the homogeneous system
calculated with $\bar{J}$. 
However, at lower hole densities where the couplings are strongly
inhomogeneous (Fig.~\ref{Fig2}a), we observe that $T_{C}$ is larger than that of the homogeneous sample. This
happens because of the competition between large (percolating)
clusters with couplings much stronger than $\bar{J}$ that tend to
increase $T_C$ and thermal fluctuations that reduce it.  In
particular, at $n_h=0.1$, $T_{C}$ happens to be close to the
mean-field result $T_{C}^{MF}= 2 \bar{J}$ (dashed line), as the result
of this competition. It is interesting to remark that this picture is
different from the pure percolation picture where thermal
fluctuations in the clusters wins and reduce $T_C$; the difference is
that the distribution is much more inhomogeneous here. We note that
our $T_C$ are much smaller than that obtained in
ref.~\cite{Salafranca}, where the same model was studied. The reason
is that here both spatial and thermal fluctuations are treated beyond mean-field virtual crystal
approximation.  

We now argue that this model gives grounds for a
Griffiths phase \cite{Salamon,Bray} above $T_C$. As discussed in
ref. \cite{Deisenhofer}, correlations in the disorder should enhance
the Griffiths phenomenon.  Indeed, it is more likely to find large
clusters with higher local ``Curie'' temperatures, as seen in
Fig.~\ref{Fig2}a.  We calculate this temperature $T_G$ from the lowest
eigenvalue $\epsilon$ of $J_{ij}$, using  $T_G=\frac{1}{3} S(S+1)|\epsilon|$ \cite{Bray}. 
Since this is a mean field estimation, $T_G$ has to be compared to $T^{MF}_{C}$.  For
$n_{h}=0.1$, we find a large $T_G = 0.11 t \approx 2.5 T^{MF}_{C}
$. On the other hand, for $n_h \sim 0.5$, the couplings are much more
homogeneous, and we have found a much smaller region for the Griffiths
phase with $T_G = 0.13 t \approx T^{MF}_{C}$. This is interesting
because it shows that $T_G$ is weakly sensitive to
$n_h$. Experimentally this phase seems to occur only in the
structurally distorted phase at low dopings \cite{Deisenhofer}, which
suggests that the origin of correlated disorder is the local
Jahn-Teller distortions, a case that is also covered by the present
model. In fact it is not clear from our study that we can exclude the
chemical origin of the correlated disorder because the Griffiths phase
shrinks as we increase the carrier density. A better treatment of
thermal fluctuations could possibly lead to the complete disappearance
of the Griffiths phase for larger dopings.

\begin{figure}[tbp]
\vspace{-1cm}
\includegraphics[width=7.0cm,angle=-90]{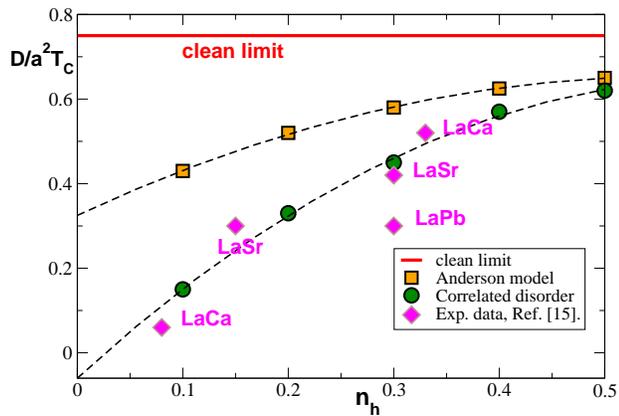}
\vspace{-0.5cm}
\caption{(Color online.) Dimensionless ratio $D/a^2 T_{C}$ of 
the spin stiffness to the Curie temperature as a function of the hole
density $n_{h}$, for the correlated and Anderson forms of
disorder. The width of the distribution of potentials was chosen to be
the same in both cases: $\epsilon_{D}=0.15 W$ ($x=0.3$) and
$\Delta=0.80W$.  Experimental results
\cite{Tapan0} are divided by the lattice constant of LaMnO$_3$ squared
($a= 3.9 \mbox{\AA}$).}
\label{Fig4}
\end{figure}

We now discuss the effect of the inhomogeneities on the long wave
length spin excitations at zero temperature. Even in the presence of
disorder, these excitations are well defined and characterized
by a spin stiffness $D$ \cite{Motomeexcitation,BouzerarDMS}, that is
calculated following \cite{BouzerarDMS}. Experiments on various manganites
show that the dimensionless ratio $D/a^{2}T_C$ ($a$ is the lattice
constant, taken to be 1 in the following) strongly varies with doping and takes values as small as
0.05 and up to 0.5 \cite{Tapan0}. This is in contrast with the
\textit{clean} double exchange model, where $D/T_C$ is a constant
equal to $0.755$ (RPA for the s.c. lattice \cite{Callen}),
independent of the hole density. We argue that the measured small values 
could be explained with the model of correlated disorder, but would
require unrealistic large strength of the disorder in the uncorrelated case.
 Fig.~\ref{Fig4} gives this ratio, calculated for both models
of disorder. Note that to allow for a direct comparison, the width of the
distributions of $\epsilon_i$ is chosen to be the same.
Close to $n_h=0.5$, $D/T_C$ does not really depend on the model,
reflecting the absence of the large scale inhomogeneities we discussed
above. When $n_h$ decreases, however, the spin stiffness is dominated by large regions of weak couplings (Fig.~\ref{Fig2}a): at $n_{h}=0.10$, $D/T_{C}$ is 3 times smaller
than that obtained with Anderson disorder. 
In order to get such small values in the Anderson disordered case, one would need a value of $\Delta$
much larger than the bandwidth, which would be difficult to reconcile with
\textit{ab initio} estimations \cite{Singh}, on one hand, and would
tend to localize all carriers \cite{Majumdar} on the other hand.  In
Fig.~\ref{Fig4} we have also compared our calculations directly with
available experimental values \cite{Tapan0}.
First we note that the overall quantitative agreement does not mean
that $\epsilon_D$ is quantitatively determined for manganites because
other interactions have been neglected here (there is anyway a
distribution of ratios for a same doping, see LaSr and LaPb in
Fig.~\ref{Fig4}, for instance which could be explained by different
$\epsilon_D$).  Nevertheless, it is interesting to see that the trend
as function of the hole density is already well captured by taking
disorder into account, and that a relatively small amount of
\textit{correlated} disorder leads to very small values of
$D/T_C$, contrary to what would be needed in the Anderson case.

To conclude, we have found that short-range correlated disorder
creates large scale spin and charge textures, particularly
inhomogeneous at low dopings. Our study suggests that describing the
disorder in a more realistic manner may be a key point to understand
experiments, as the occurence of a Griffiths phase
observed for low dopings. In addition, we have found that correlations
in the disorder tend to increase $T_C$ because of the presence of
large clusters of strong couplings, and decrease the spin stiffness
$D$. This results in very small ratios $D/T_C$, consistent with what
has been measured in manganites but hardly compatible with Anderson
disorder. The dimensionless ratio $D/T_C$ appears as a good measure of
the inhomogeneous character of the magnetic state, a conclusion that
may apply beyond manganese oxides.

We thank M. Clusel, M. Hennion, P. Majumdar, Y. Motome, F. Moussa and S. Petit for stimulating discussions. O.~C. thanks the ILL for hospitality.

\end{document}